# FAULT TOLERANT WIRELESS SENSOR MAC PROTOCOL FOR EFFICIENT COLLISION AVOIDANCE

Abhishek Samanta[1] and Dripto Bakshi[2]

[1]Department of Computer Science and Engineering, Jadavpur University, Kolkata, India
avisheksamantajunior@gmail.com
[2]Department of Computer Science and Engineering, Jadavpur University, Kolkata, India
bakshi.dripto@gmail.com

## ABSTRACT

*In sensor networks communication by broadcast methods involves many hazards, especially collision. Several MAC layer protocols have been proposed to resolve the problem of collision namely ARBP, where the best achieved success rate is 90%. We hereby propose a MAC protocol which achieves a greater success rate (Success rate is defined as the percentage of delivered packets at the source reaching the destination successfully) by reducing the number of collisions, but by trading off the average propagation delay of transmission. Our proposed protocols are also shown to be more energy efficient in terms of energy dissipation per message delivery, compared to the currently existing protocol.*

## KEYWORDS

Wireless Sensor Networks, Energy efficient, Data propagation, Single path, Multi path, Collision avoidance, Success rate

## 1. INTRODUCTION

Sensor networks are particularly useful in collecting data from inaccessible terrains which serve various purposes in further investigation. Sensor network comprises of a large number of nodes (often termed as motes) which are randomly distributed very densely in the area concerned. The data collected by each node is transmitted to subsequent nodes and thus finally resulting in the reporting of the data to the sink which can be considered as a destination for delivering data.

### 1.1. Motivation

The common methods employed to resolve the MAC layer problems cannot be employed in sensor networks. CSMA cannot be employed [1] because the csma-based protocols involve broadcasting of control messages between the sender and receiver, in order to enable the sender to acquire the transmission media, and these eventual broadcasts result in a collision between these control messages.

#### 1.1.1. Existing back off scheme

In order to resolve the MAC layer problem of collision a few back off schemes have been designed, namely SRBP (Simple Random Back Off Protocol), ARBP (Adaptive Random Back Off Protocol) and RARBP (Range Adaptive Random Back Off Protocol) [2]. In the back off protocols the broadcasts are delayed by a certain back off period, i.e. the nodes assume a random back off time and then transmit the data packet it wants to transmit. Thus simultaneous broadcasts and consequent interference of data resulting in collision is prevented, since the broadcasts of the several nodes are spread over time.

The backs off schemes when applied on the AODV protocol significantly improve the success rate (which is defined as the percentage of broadcasted packets reaching the final destination). Amongst all the back off schemes used till date ARBP achieves the highest success rate (i.e. success rate of 90%). We hereby propose a back off scheme which achieves a success rate of around 95%, which is significantly higher than those achieved by the existing protocols. The back off schemes are adopted to prevent collisions by spreading the broadcasts over time. Since the nodes adopt different back off periods before transmitting a data packet, simultaneous broadcast is prevented and hence reducing the number of collisions. The continuous set from which the back offs are selected ($T_{min}$, $T_{max}$) varies from one protocol to another [2].

But these protocols ignore the fact that more than one node may select the same back off period or any node can chose a back off period in such a way that it starts to transmit its packet of data before another previously transmitted data packet has finished transmission, thereby resulting in collisions in both cases. This problem has been approached from two directions. In one of our schemes, it has been ensured that nodes always choose a back off period in such a way so as to overcome the above mentioned deficiencies. In other scheme, the backflow of packets (the flow of packets in the direction opposite to the direction of gateway node, i.e. sink) has been controlled. Both of these schemes ensure the low incidence of collisions and thereby increase the success rate of overall protocol.

## 2. PROPOSED MAC PROTOCOL

### 2.1. Informed Back Off selection Protocol (IBSP)

The protocol primarily consists of two phases. During the first phase the back off period selected by a node for data transmission is informed to the neighbouring nodes in the radius of transmission of that node, and thus the other nodes select their respective back off periods accordingly. Thus this internodal sharing of information about the back off periods chosen during data packet transmission, further eliminates the possibility of coincident transmission, and thereby reduced collisions. During the second phase actual data transmission takes place.

#### 2.1.1. First Phase

In the first phase when a node is ready to broadcast a data packet, it first sends a control packet consisting of a MAC header in which the back off period for data transmission is appended. The back off periods used for the transmission of control packets is selected from the continuous set ($T_{min}$, $T_{man}$). Here $T_{min}$ is the minimum time taken for the node to node transfer of packets and $T_{max}$ is calculated according to following subroutines.

##### 2.1.1.1. The density-sensing subroutine ($P_{density}$)

We call $d_l$ the average number of neighbours a node senses over a certain area (i.e. the local density). Initially $d_l=d_i$, where $d_i$ is set to reflect the (expected) conditions of the network; the expected number of neighbours of a node is related to the network density, i.e. $d_i \alpha d$. $P_{density}$ maintains a table where it stores all the senders' ids encountered along with a time counter indicating the time the message was received. In fact, the subroutine is continuously inspecting all packets received and updates the local table. For every entry in the table a counter is defined that is initialized to a predefined period of time called $t_{inactive}$ (e.g. $t_{inactive}$=1hour). Periodically, $P_{density}$ will go through the list of ids and remove those neighbours whose counter has expired. Therefore, the $P_{density}$ table needs $o(d_i)$ entries.

##### 2.1.1.2. The message traffic sensing subroutine ($P_{traffic}$)

We call $I_l$ the average number of distinct messages received per period of time by a node (i.e. the local message traffic). Initially $I_l=I_i$, where $I_i$ is set to reflect the (expected) conditions of the network. In fact, the expected message traffic handled by a node is related to the global message traffic i.e. $I_i \alpha I$. $P_{traffic}$ maintains a variable that is used to count the total number of messages received within each time period (i.e. every 1sec the counter is reset). Every message received is considered in the calculation of $I_l$ besides the fact that in multi-path propagation, several messages carrying duplicate data might exist. Its application is specific to suppress duplicate messages. Remark that the calculation of $d_l$ and $I_l$ is performed dynamically and is subject to change over time; in this case, notion of $d_l(t)$ and $I_l(t)$ is introduced where $d_l(0)=d_i$ and $I_l(0)=I_i$. We fix the value of $T_{min}$ to a value little higher than the amount of time it takes to transmit a packet. On the other hand, the value of $T_{max}$ will change once the network is running, and for this reason again we use the notion of $T_{max}(t)$. There are several ways to calculate a new value for $T_{max}$; we here present the change of $T_{max}$ as a function of successive measurements of a local density, the local message traffic and the current value of $T_{max}$:

$$C_d(t) = T_{max}(t-1) \times ((d_l(t) - d_l(t-1)) \div (d_l(t) + d_l(t-1))) \quad (1)$$

$$C_t(t) = T_{max}(t-1) \times ((I_l(t) - I_l(t-1)) \div (I_l(t) + I_l(t-1))) \quad (2)$$

Then based on $C_d(t)$ ad $C_t(t)$, $T_{max}(t)$ is calculated as follows:

$$T_{max}(t) = T_{max}(t-1) + \alpha \times C_d(t) + \beta \times C_t(t) \quad (3)$$

Where $\alpha \in [0, 1]$ and $\beta \in [0, 1]$, two parameters that adjust the way $T_{max}(t)$ is calculated and its dependency on the two components. Fine grained adjustment can be achieved for different kinds of application requirements. For applications with constant data generation rate, by setting $\beta=0$ the dependence on traffic rate is totally removed and IBSP adapts only based on the sensed local density change. Furthermore, $\alpha$ and $\beta$ determine how drastically the protocol reacts to changes. For values close to 0 little adjustment is made, while values close to 1 allow more drastic adaptation.

Selection of back off for data transmission takes place according to the following scheme. Whenever a node transmits the control packet to its neighbouring nodes, the nodes on receiving the control packet, reads the back off selected by the sender (say T), and store it in their respective memories. Then each of the receiving nodes mark the zone $(T-T_{min}, T+T_{min})$ as forbidden zone, i.e. no back off further selected will lie in this zone. In this way the next node (whose timer expires next) selects a back off for data transmission and informs it to the other neighbouring nodes, via the transmission of control packets. In this way, by sharing of control packets, the subsequent sharing of information about the back off time selected by the nodes takes place. The back off periods for data transmissions are selected from the continuous set (Tmin+Tmax, 2Tmax). This set is selected because of the following reason. During the control packet transmission, as mentioned earlier, the maximum back off that can be selected by any node is equal to $T_{max}$. Now since the maximum time for node to node transmission is $T_{min}$, so the transmission of control packets must be over by the time $(T_{max}+T_{min})$. Again since the data packet transmission also follows the scheme stated above, the back off period selection must take place from a continuous set having a width of $(T_{max}-T_{min})$. Thus the back off period selection for data transmission takes place from the set $(T_{max}+T_{min}, T_{max}+T_{min}+T_{max}-T_{min})$ i.e. $(T_{max}+T_{min}, 2T_{max})$.

The node receiving the data packet first is the first node to let its neighbouring nodes know about the back off, that it is going to take during data propagation in second phase, selected by it. Now when the first node informs all the other nodes in its neighbourhood (i.e. in range of radio transmission) about the back off period it has selected for phase II, i.e. the data packet transmission phase, (say T), then all the informed nodes mark $(T+T_{min}, T-T_{min})$ as the forbidden zone while choosing their back offs, because if any node selects a back off in the above mentioned zone then a collision is bound to occur with the packet delivered by the node which has chosen a back off of T.

Now again the second node chooses a back off period from the set ($T_{max}+T_{min}, 2T_{max}$) apart from the forbidden zone created by the first. The back off thus selected by the second node is informed to all the nodes in its range of radio transmission. When the third node chooses its back off for data packet transmission, it chooses the back off from the set ($T_{max}+T_{min}, 2T_{max}$) apart from the forbidden zones specified by the first two nodes. In this way, any node chooses its back off for data packet transmission phase from the set ($T_{max}+T_{min}, 2T_{max}$) apart from the forbidden zones created by the nodes preceding it.

Thus for nodes in the range of transmission of each other, there is the sharing of information regarding the back off periods selected for the data transmission of phase II and thus collisions is minimum.

We have mentioned earlier that the information sharing about chosen back off and hence careful subsequent back off selection amongst the nodes within each other's range of radio transmission. Let us say that these nodes form a group. Now, all the nodes in a particular group are well aware of the back offs chosen by the other nodes of that group and accordingly have selected their own back offs so as to avoid collision. Now let us consider that a set of nodes $N_i$ be part of two groups $G_1$ and $G_2$. Now there is sharing of information regarding back off selection in group $G_1$ as well as in group $G_2$. But since these two groups share few nodes between them, so there will also be an intra group sharing of information. Here the common nodes serve as the link between the two individually coordinated groups, and this leads to a unified coordination amongst all nodes of groups $G_1$ and $G_2$. Since nodes are more or less always shared by two or more groups so the intra group sharing of information regarding the chosen back off selection eventually leads to an inter group coordination as well. In other words coordination is achieved amongst all nodes trying to send data packets at any instant of time, i.e. the back offs are so selected and coordinated that the number of collisions is minimum.

### 2.1.2. Second Phase

In this phase the actual data transmission takes place. The nodes assume the back offs previously chosen by them in phase I, and eventually deliver the data packets, thus reducing the number of collision to the minimum.

### 2.1.3. Mathematical Explanation

In this mathematical explanation we try to establish relationship between success rate of IBSP and that of ARBP.

In the ARBP protocol the back off period is selected from the interval ($T_{min}, T_{max}$). $T_{min}$ is assumed to be time required for node to node transmission.

In our proposed protocol in the first phase, selection of back off, back off is selected from the interval ($T_{min}, T_{max}$) and in the second phase from the interval ($T_{max}+T_{min}, 2T_{max}$). In both cases the width of the interval is ($T_{max}-T_{min}$).

Let us consider a situation where N number of nodes are uniformly distributed over a region of radius R and r be the range of radio transmission [Fig.1]. So, the number of nodes present in range of a node is as follows,

$$n = (r^2 \div R^2) \times N \qquad (4)$$

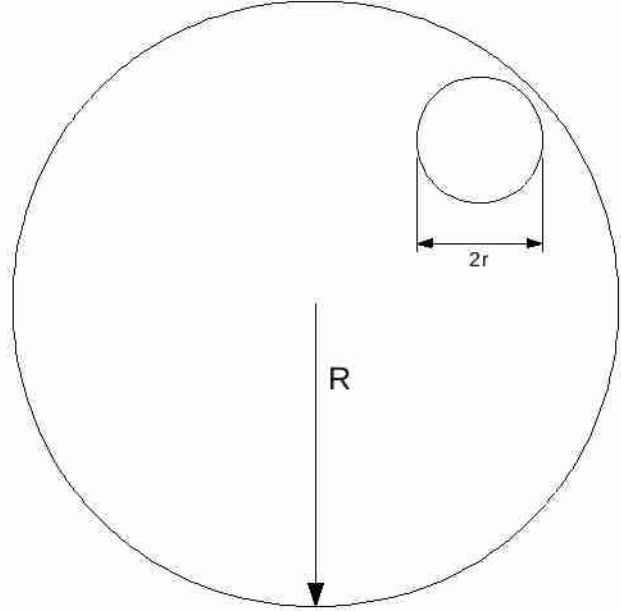

Fig 1

Let us consider that a random node is transmitting data following ARBP protocol. In order to ensure that no collision occurs during its transmission the back off selected should not fall in the range $(t-T_{min}, t+T_{min})$ where t is the back off selected by any random neighbour, i.e. any back off selected from this time zone of span $2T_{min}$ results in collision of packet. There will be n such neighbours. For n distinct such set, the forbidden time zone is $2nT_{min}$. So, probability of resulting in collision is as follows,

$$Prob_n = (2 \times n \times T_{min}) \div (T_{max} - T_{min}) \quad (5)$$

In case of IBSP, we have to consider the two phases distinctly.

Phase I: Probability of occurrence of collision is same as that in ARBP, ie $Prob_n$

Phase II: For a node, total forbidden interval = $2nT_{min}$. For a node which has not known a back off for phase II due to collision in phase I, probability of collision is,

$$Prob_{phase\ II} = (2 \times T_{min} \times n) \div (T_{max} - T_{min}) \quad (6)$$

Hence probability of collision of data packets in IBSP is,

$$Prob_{IBSP} = Prob_{phase\ I} \times Prob_{phase\ II}$$

$$= ((2 \times n \times T_{min}) \div (T_{max} - T_{min}))^2$$

$$= (Prob_{ARBP})^2 \quad (7)$$

Since, $Prob_{ARBP} < 1$, so $Prob_{IBSP} < Prob_{ARBP}$, which means that collisions among packets with IBSP are less probable to occur than with ARBP.

**2.1.4. Algorithm of IBSP**

a. $T_{min}$ → minimum time taken for node to node delivery

b. $T_{max}$ → maximum back off time optimized in order to reduce the number of collisions and is calculated according to ARBP scheme.

c. eligible_set → a set from which to select a back off from during data transmission phase (Phase II).

d. queue_packet → queue of packets need to be broad casted out.

Phase I:

　a.　eligible_set ← $[T_{max} + T_{min}, 2T_{min}]$

　b.　backoff_phase_I ← k ϵ $[T_{min}, T_{max}]$ according to ARBP

　c.　backoff_phase_II ← t ϵ eligible set according to ARBP

　d.　Wait for backoff_phase_I

　　　**if** no packet received **then**

　　　　Send backoff_phase_II to all_neighbours

　　　**else**

　　　　**if** the packet is a data packet **then**

　　　　　enqueue(neighbour_packet to queue_packet)

　　　　**else**

　　　　　$t_n$ ← back off selected by neighbouring node for phase II

　　　　**end if**

　　　　eligible_set ← eligible_set - $[t_n - T_{min}, t_n + T_{min}]$

　　　　**goto** step b

　　　**end if**

Phase II:

　a.　wait for backoff_phase_II

　b.　packet_data ← dequeue(queue_packet)

　c.　broadcast(packet_data)

　d.　**if** data packet from neighbour received **then**

　　　　neighbour_packet_hopcount ← hop_count_level_of_node

　　　　enqueue(neighbour_packet to queue_packet)

　　　**end if**

**2.2. Direction Aware IBSP (DAIBSP)**

This protocol is more like a paradigm which can be incorporated on the top of any multicast wireless sensor MAC protocol. The whole life cycle of DAIBSP can be classified into two types of packet transmissions and each packet is transmitted according to the rule specified by the underlying MAC protocol. One type of transmissions consists of only beacon packets and are allowed to be sent out in the opposite direction of sink. In the second type, transmissions take place with the actual data packet and other ancillary packets that need to be sent out following the underlying MAC protocol, which in this case is IBSP.

### 2.2.1. Beacon Packet Transmission Phase

When the topology is ready to be used, the sink node broadcasts a beacon packet consisting only of a field which keeps record of hop count and this beacon packet is rebroadcasted by each node to their neighbours. When a node receives the beacon packet, it compares its present hop count level (which is initially set to -1, which indicates that the node has not received a packet yet) with the hop count field in the beacon packet; if the hop count field value in the beacon packet is greater than or equal to present hop count level of the node then this information is discarded. Else the hop count level of the receiving node is calculated according to eq.8 and the beacon packet is rebroadcasted with hop count value of the node. In this way the beacon packet is sent out from the centre to the peripheral regions, restricting any inbound movement of the packet.

$$HopCount_{node} = HopCount_{inCommingPacket} + 1 \qquad (8)$$

### 2.2.2. Data Packet Transmission Phase

Every node maintains a table consisting of two fields, viz. hop count and a timer. At any moment the hop count of a node is associated with the minimum entry from the hop count table with live timer. While transmitting a packet the hop count of the node is put in the MAC header of the packet. It is obvious that all the nodes are not possible to be informed about their hop count in phase I. For those nodes, i.e. for nodes which do not have any knowledge about their position in the topology can only listen to the message traffic around it till it is assigned to some value. When a node gets an incoming packet it looks at the hop count information of the packet. If the incoming hop count is greater than or equal to that of the node, it would be put in the table with a live counter. Since initially all the nodes are assigned to a very small hop count value, i.e. -1, in this way the nodes that were not informed in the phase I would get information about their position. In this phase when a node gets a packet from a node whose hop count value is lesser, the packet is dropped and is not retransmitted.

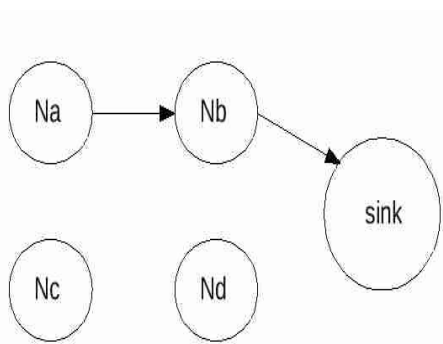
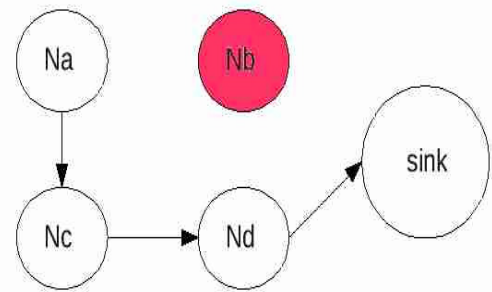

          Fig. 2                                        Fig. 3

In this way the packets always are forwarded from the source to the sink and hence the undesirable backflow (the flow of data in the opposite to the source-sink direction) of information is hindered. Thus the extra consumption of energy that results from the subsequent collisions due to the above mentioned backflow is avoided.

The hop count level table is maintained to protect the network from debacle when some unforeseen catastrophe happens. Let us consider 3nodes, $N_a$, $N_b$, $N_c$. $N_a$ is at a distance $d_b$ from the sink through $N_b$ and at a distance $d_c$ through the node $N_c$, where $d_b < d_c$. So the hop count of $N_a$ is $d_b$ [Fig. 2]. Now if $N_b$ dies, the packets emanated from $N_a$ is dropped by $N_c$ since the hop count level in the packets is less than that of $N_c$. This happens only till the timer associated with $d_b$ is alive. After the time associated with $d_b$ expires, $N_a$ assigns $d_c$ as its hop count and then the data transmission is again restored [Fig. 3].

### 2.2.3. Mathematical Model

Let us consider an area with radius R where N no. Of nodes have been uniformly deployed and the gateway node has been kept at the centre of the circular area [Fig. 4]. Now, in this mathematical model a comparison of number of transmission taken place for the IBSP and DAIBSP is made.

Let us consider an ideal environment where no collision among packets takes place. The only condition that restrains a perpetual transmission is the non-cyclical nature of path of packets. That means a packet cannot be handled by a node twice.

For the IBSP, since nodes do not have any knowledge about their position in the topology, the packets are sent out in all directions. Since the nodes are uniformly distributed, the number of nodes present in a particular region is directly proportional to the area. So, if the occurrences of events are perfectly random, the probability of a node at a distance r from sink sensing an event is,

$$\text{Prob}_r = 2\pi r dr \div \pi r^2 \qquad (9)$$

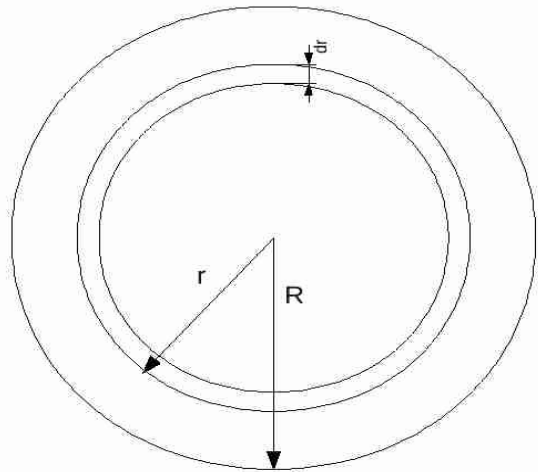

Fig. 4

Since a packet is sent out in all the directions and the packet cannot follow a cyclic path, for a data packet sent out by a node there can be N number of subsequent transmissions. So, expected number of transmission for a sensed data packet (original packet) is,

$$E(r) = \text{Prob}_r \times N = (2\pi r dr \div \pi r^2) \times N \qquad (10)$$

So, according to ARBP, the expected amount of transmissions taking place for generation of an event is as follows,

$$E_{xmission} = {}^R\!\int_0 E(r) = {}^R\!\int_0 (2\pi r dr \div \pi r^2) \times N = N \qquad (11)$$

For DAIBSP, probability of getting an event sensed by a node at a distance of r is as already stated, $\text{Prob}_r$. In this protocol, since the nodes are aware of their position in topology, only the nodes inside the circle of radius r participate in subsequent packet transmission, when a node at a distance of r transmits. Since cyclical paths are not allowed, each node can forward a packet only once. So number of transmission taking place for a transmission by the node at distance r is,

$$N_{nodes} = (\pi r^2 \div \pi R^2) \times N = (r^2 \div R^2) \times N \quad (12)$$

So, expected number of subsequent transmissions taking place by a sensed data packet transmission is,

$$^1E_{xmission} = {}^R\!\int_0 (Prob_r \times N_{nodes})$$
$$= {}^R\!\int_0 ((2\pi r dr \div \pi r^2) \times ((r^2 \div R^2) \times N))$$
$$= N \div 2 \quad (13)$$

So, from eq.11 and eq.13 it can be stated that number of transmissions are almost halved by introducing DAIBSP over IBSP, which effectively conserves energy and increases the longevity of the nodes.

### 2.2.4. Algorithm of DAIBSP

a. pkt ← packet from neighbour
b. enlist(pkt,t) ← put (hop count in pkt)+1 associating it to a timer with value t in the
   hop count table
c. eligible_set ← [$T_{max} + T_{min}$, $2T_{max}$]
d. backoff_phase_I ← k ϵ [$T_{min}$, $T_{max}$] according to ARBP
e. backoff_phase_II ← t ϵ eligible set according to ARBP
f. queue_packet ← queue of packets need to be broadcasted out

Building Phase:
receive(sink_packet)
**if** hop_count_of_node ≤ sink_packet_hop_count **then**
    hop_count_level_of_node = sink_packet_hop_count
    sink_packet_hop_count ← sink_packet_hop_count + 1
    broadcast(sink_packet_hop_count)
**end if**
**loop**
  Phase I:
    Wait for backoff_phase_I
    **If** no packet received **then**
      Send backoff_phase_II to all neighbours
    **else**
      enlist(pkt,t)
     **if** hop_count_of_node > hop_count_in_pkt **then**
       hop_count_of_node = hop_count_in_pkt + 1
     **end if**
     **if** the packet is a data packet **then**
       enqueue(neighbour_packet to queue_packet)
     **else**
       tn ← bacloff selected by neighbouring node for phase II
     **end if**
     eligible_set ← eligible_set − [tn − Tmin, tn + Tmin]
     goto step d
    **end if**

Phase II:
   wait for backoff_phase_II
   packet_data ← deque(queue_packet)
   broadcast(packet_data)
   **if** data packet from neighbour received **then**
     enlist(pkt,t)
     **if** hop_count_of_node > hop_count_in_pkt **then**
       hop_count_of_node = hop_count_in_pkt + 1
     **end if**
     **if** the packet is a data packet **then**
       enque(neighbour_packet to queue_packet)
     **end if**
   **end if**
**end loop**

## 3. EXPERIMENTAL EVALUATION

To evaluate the performance of the proposed protocol we have done simulation analysis in NS2 (network simulator). We start out experiment by dropping n ϵ [500, 1500] within a smart dust plane of 500m × 500m in dimension. We also generated 2000 events by randomly selecting a particle in the network for each event. With event generation rates (λ) 5, 8, 10. For DAIBSP a table with 10 least hop counts were maintained with a initial timer value of 2sec. The transmission range for one node has been set to 50m. We here define the success rate as,

$$\text{SuccessRate} = (\text{DataReceived} \div \text{DataSent}) \times 100\%$$

From our first experiment we have found that the success rate for a sensor network increases about 4% when we use IBSP instead of ARBP and DAIBSP increases success rate by 1% more on IBSP protocol. The success rate increases drastically as the node density goes from 500 to 1000 and decreases by little bit as the nodes density increase further. With node density 1000, the success rate reaches around 95%. The result remains almost same when simulated with λ = 8. The average delay, on the other hand, increase by 80% over normal ARBP. With lower node density the success rate of IBSP protocol was almost equal to that of ARBP. This can be explained by the fact that, at lower node density the expectation of two nodes being in the range of each other, is less than that in case of higher concentration of nodes. So at n = 500 the lower amount of success rate is mainly because of the unreachability of nodes rather than the collision of packets. But as the concentration of the nodes gets increased the reachability problem is overtaken by the problem generated by the overcrowded traffic which causes packets to drop.

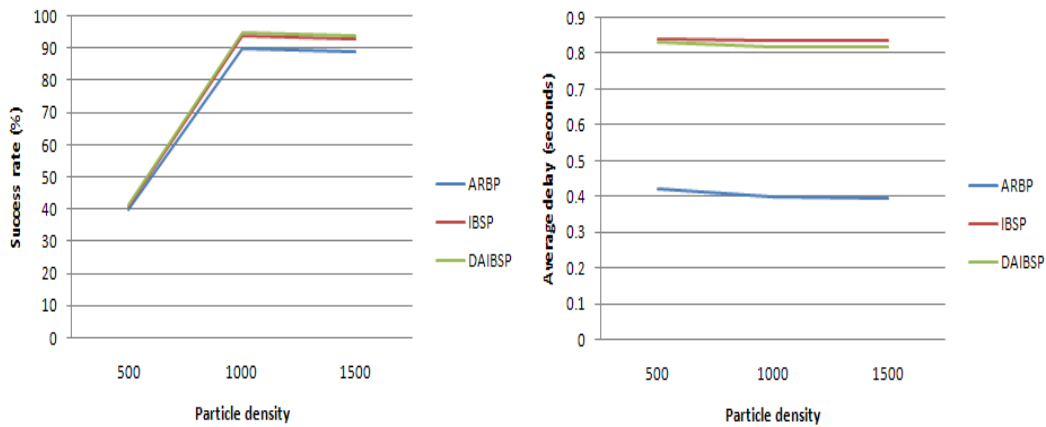

Fig.5: Success rate and average delay for varying node densities (n ϵ [500, 1500]) with λ=5, α=1 and β=0

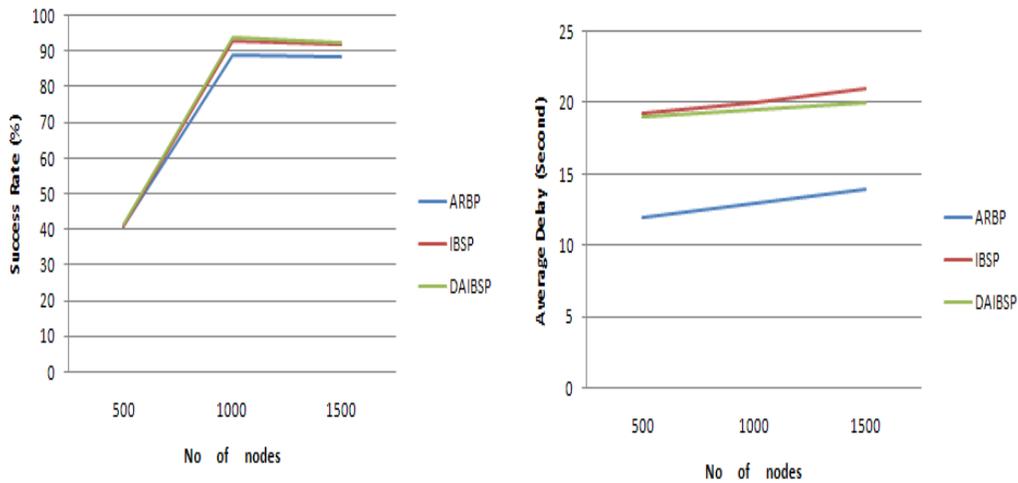

Fig 6: Success rate and Average delay for varying node densities (n ϵ [500, 1500]) with λ=8, α=1 and β=0

This is where IBSP performs better than ARBP. But as the n increases to 1500, the message traffic becomes so high that the beacon packets from the nodes starts colliding more in the first phase of information sharing, due to which the neighbours of a node do not get properly informed about the back off taken by the node. Due to this miscommunication the success rate of IBSP decreases to around 90%. But both in ARBP as well as in IBSP the data packet can go anywhere in the network irrespective of the direction of the sink, as long as the packet is not moving in a circular path. These adhoc movements of data packets increase probability of two packets getting collided. To avoid this, DAIBSP only allows movement of packets in the direction of sink. This scheme ensures that a packet is confined within the area having radius equal to the hop count of the node where the packets have been emanated from. Due to this reason the success rate of DAIBSP is even greater than IBSP and since it does not take any extra transmission on simple IBSP to share information, the average delay remained almost

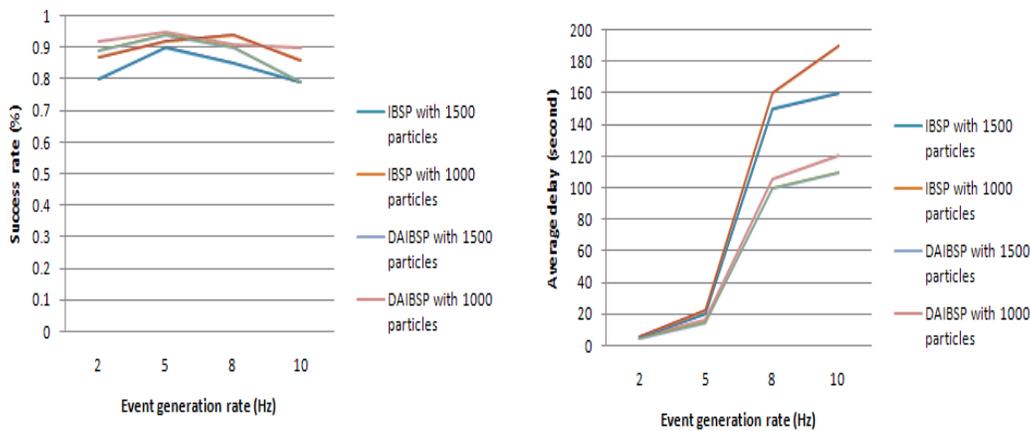

Fig 7: Success rate and average delay for IBSP and DAIBSP with α = 1 and β = 1 for variable event generation rate (λ ϵ [2, 10]) and different particle densities (nϵ[1000,1500])

same. Simple IBSP scheme observes larger packet drops [Fig. 10] because phase I of the scheme when no effective information sharing takes place, follows the ARBP scheme, which inherently has 90% success rate. In phase II, when actual data transmission takes place, some of the data packets still manage to have collisions. So for every data packet transmission the number of colliding packets gets increased. But since DAIBSP inhibits backflow of information, the number of colliding packets drastically falls down.

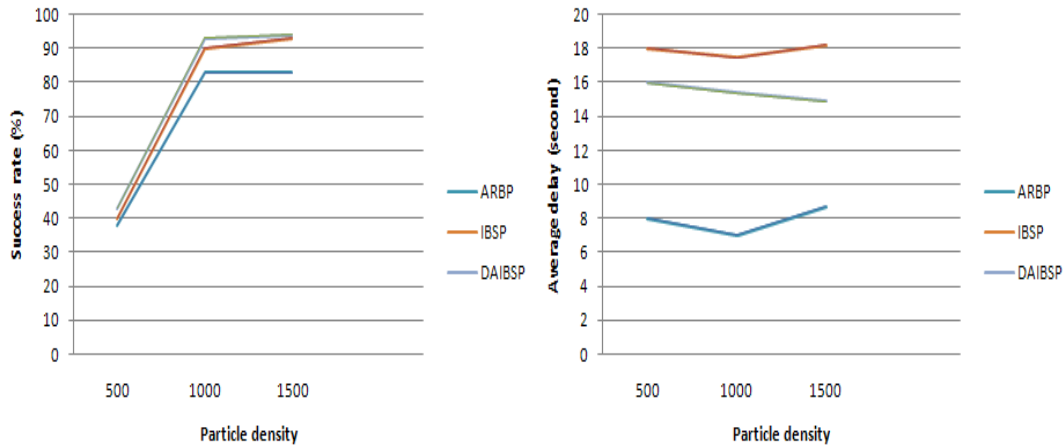

Fig 8: Success rate and Average delay for varying node densities (n ϵ [500, 1500]) with λ=5, α=1 and β=1

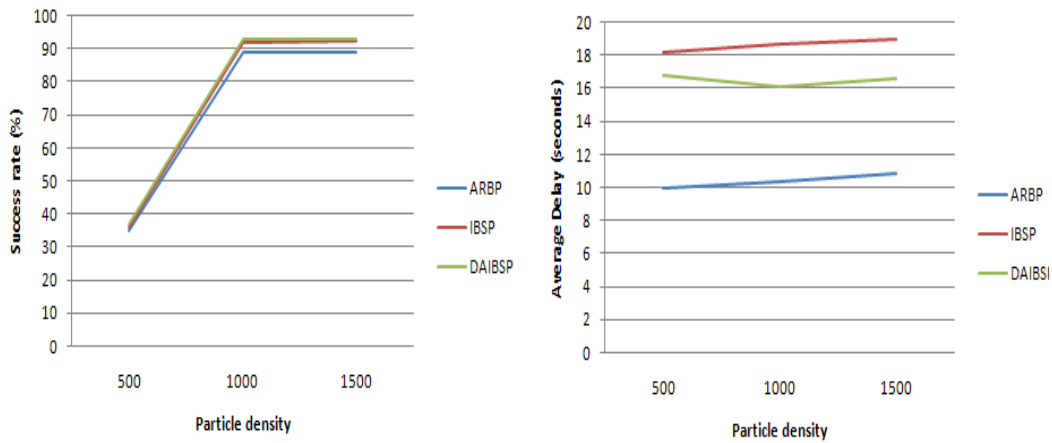

Fig 9: Success rate and Average delay for varying node densities (n ϵ [500, 1500]) with λ=5, α=0 and β=1

In Fig.7, some interesting features can be seen. As event generation rate increases from 2 to 5, the success rate increase for IBSP but they remain almost same for DAIBP. One reason for this observation might be due to the fact that subroutine $P_{density}$ and $P_{traffic}$ largely depends on the hardware being used in the physical layer. Since no active communication is established among the neighbouring nodes for density sensing or message traffic sensing protocol, so sensing protocols largely depend on the packets broadcasted during the data transmission. If occurrence of an event is not very frequent, that leads to unsensed neighbouring nodes. This causes erroneous calculation of back off and which in turn clamps down the success rate. As the event generation rate increases, the overall message traffic also increases accordingly, which makes nodes sense their surroundings more precisely and so success rate increases. For DAIBSP, even before an event is sensed, a beacon packet is transmitted by the sink and this packet is distributed throughout the whole topology and it itself generates lots of packet transmission. This initial packet transmission causes a better sensing of surroundings by nodes which directly affects the success rate. For this reason the success rate for DAIBSP remains almost same for message generation 2 and 5. As the event generation rate is increased further the success rate remain almost same for IBSP and DAIBSP. But a strange observation takes place for particle n = 1000 and λ = 8. Under this situation success rate for IBSP overtakes that of DAIBSP. This is attributed to the fact that, as λ is increased, more collision occurs in the first phase of information sharing of IBSP and DAIBSP. So, the probability of collision of packets increases abruptly. But IBSP includes a larger set of nodes for data transmission, so more paths consisting of nodes further away from sink are considered for packet forwarding. But DAIBSP considered lesser number of nodes only by taking into account the nodes which are at a distance less than or equal to the distance of the source node. So, for this there is no possibility for selecting a path containing nodes far away. This inherent problem of DAIBSP causes its success rate become less than that of IBSP. But with node density 1500, packet transmissions increase so largely that the finding of paths with faraway nodes does not help IBSP much and this causes its success rate to come down below DAIBSP. As event generation rate is increased further to 10, collision among the packets during the phase I of IBSP and DAIBSP causes the data packets to collide during the phase II of transmission thereby decreasing the success rate further.

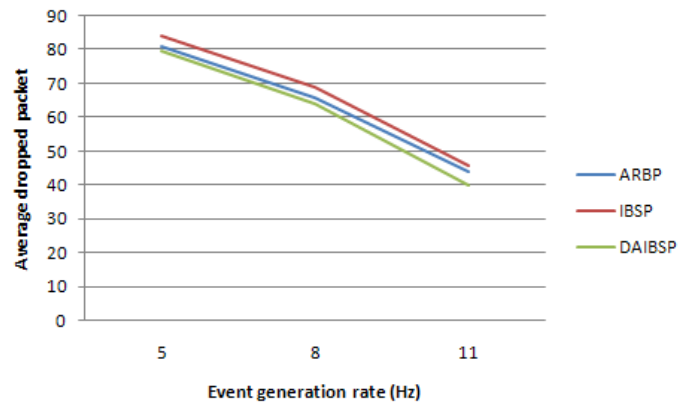

Fig 10: Average number of dropped packets over event generation rate with n = 1000 for α = 1 and β = 0

From the average delay curve of Fig. 7, it can be seen that from λ = 2 to λ = 5 the delay increases fairly slowly. But from λ = 5 to λ = 8 there exists a sharp increase in delay. This is caused by the excessive amount of collisions taken place for λ = 8. Due to this increased number of collisions packets take longer path to reach sink, which causes large amount of delay seen.

With α = 1 and β = 1 the dependence on $C_d(t)$ increases which clearly indicates greater dependency on message traffic. With greater message traffic the chances of collision in ARBP increases. In IBSP and DAIBSP the efficiency factor regarding collision avoidance as compared to ARBP becomes more pronounced in this simulation. This greater improvement in efficiency when message traffic factor is taken into consideration is due to the informed back off approach leads to this multiplier effect.

It can be seen that among Fig 5, 8 and 9, the average delays are minimum for α = 1 and β = 1. This is because the fact that there are multiple copies of a single data packet present at a single point of time, in this topology. So, although one such packets collides leading to an unsuccessful transmission, but another image of the packet can take longer paths to finally reach the gateway node. This would directly affect the average delay of the transmission. Keeping this in mind, it can be assumed that a rough reciprocal relationship exists between success rate and average delay, i.e. more the success rate smaller is the average delay which has been reflected in the graphs.

## 4. CONCLUSION

The proposed MAC protocol further resolves the problem of collisions that is inherent to broadcasting in sensor networks. In our protocol we tried to incorporate some techniques which were previously used in higher levels in network stack, in MAC layer. Our protocol is first of its kind to make MAC layer handle the hop count information and there by incorporate source-sink direction sensing capability in broadcasting protocols. The protocol reduces collision to a great degree (95% success rate approx) which is a clear improvement over the existing collision resolving protocols (90% success rate the best achieved reliability till date). The reduced number of collisions leads to greater energy efficiency by reducing the number of dropped packets, with trade off with the average delay in transmissions.

Reducing the time delay trade off calls for future deliberation.

## REFERENCES

[1] Liker Demirkol, Cem Ersoy, Fatih Alagoz, MAC protocol for wireless sensor networks: a survey IEEE Communications Magazine, vol 44, iss. 4, pp. 115-121, 2006..

[2] Ioannis Chatzigiannakis, Athanassios Kinalis, Sotiris Nikoletseus Wireless Sensor Network Protocols for efficient collision avoidance is Multipath Data Propagation in Proceedings of the 1st ACM international workshop on Performance evaluation of wireless ad hoc, sensor and ubiquitous networks, 2004..

[3] A. Boukerche, R. Pazzi and R. Araujo A novel fault tolerant and energy aware based algorithm for wireless sensor network in Algorithmic Aspects of Wireless Sensor Networks, 2004.

[4] Chritos H. Papadimitriou* and David Rataczak On a Conjecture Related to Geometric Routing in algorithmin aspect of wireless sensor network, 2004.

[5] AN-SWOL HU and SERGIO D. SERVETTO Algorithmic aspects of the time synchronization problem in large-scale sensor network in Mobile Networks and Application, 2005.

[6] Tijs van Dam, Keon Langendoen An Adaptive Energy-Efficient MAC Protocol for Wireless Sensor Networks in The First ACM Conference on Embedded Networked Sensor System (SenSys 2003).

[7] Keon Langendoen, Niels Reijers Distributed localization in wireless sensor networks: a quantitative comparison in Computer Networks (Elsvier), apecial issue on Wireless Sensor Networks, November 2003.

[8] Seungkyu Bac, Dongho Kwak, Cheeha Kim Traffic-Aware MAC Protocol Using Adaptive Duty Cycle for Wireless Sensor Networks in the International Conference on Information Networking, 2007.

[9] Sangik Cha, Jaehoon Ko, Soonmok Kwon and Cheeha Kim 3-hop Ahead Path Reservation Scheme for Expedite Traffic in Wireless Sensor Networks in International Conference on Information Networking (ICOIN), 2009.

[10] The network simulator ns2. http://www.isi.edu/nsnam/ns.